\documentclass[prb,twocolumn,floats,superscriptaddress,showpacs,floatfix]{revtex4}
\usepackage{graphicx} 

\begin{document}

\title{Nonlocality and fluctuations near the optical analog of a sonic horizon}
\author{S. Bar-Ad},
\affiliation{
Raymond and Beverly Sackler Faculty of Exact Sciences,\\
School of Physics and Astronomy, Tel-Aviv University, Tel-Aviv
69978, Israel }
\author{R. Schilling}
\affiliation{Johannes Gutenberg University, Mainz, Germany}
\author{V. Fleurov} \affiliation{
Raymond and Beverly Sackler Faculty of Exact Sciences,\\
School of Physics and Astronomy, Tel-Aviv University, Tel-Aviv
69978, Israel }
\begin{abstract}
We consider the behavior of fluctuations near the sonic horizon and
the role of the nonlocality of interaction (nonlinearity) on their
regularization. The nonlocality dominates if its characteristic
length scale is larger than the regularization length. The influence
of nonlocality may be important in the current experiments on the
transonic flow in Kerr nonlinear media. Experimental conditions,
under which the observation of straddled fluctuations can be
observed, are discussed.
\end{abstract}

\pacs{42.65.-k, 47.40.Hg, 47.60.Kz, 04.70.-s}
\maketitle


\section{Introduction.}

A possibility of creating a sonic horizon in a transonically
accelerating flow \cite{U81}, mimicking either black or white hole
horizons, inspired proposals and attempts to realize them. Quite a
number of theoretical models realizing such artificial black (white)
holes has been put forward
\cite{R00,G05,BLV03,CFRBF08,RPC09,NBRB09}. Experimentally a
white-hole horizon was observed in optical fibers,\cite{PKRHK08}
where the probe light was back-reflected from a moving soliton, and
a black-hole horizon was observed in a Bose-Einstein condensate
(BEC) system \cite{LIBGRZS10}. A radiation in optical fibers was
observed \cite{BCCGORRSF10}, which is a promising contender to being
analogous to Hawking radiation in a table-top experiment. A "horizon
physics" is currently intensively studied also in the surface water
waves \cite{RMMPL08,RMMPL10,WTPUL11}. The whole field of analog
gravity has been recently reviewed in Ref. \onlinecite{R12}.

Propagation of the coherent laser beam in Kerr nonlinear media is in
its many aspects analogous to a flow of fluid. Properties of such
"luminous fluid", in particular formation of the so called
"dispersive shock waves" were studied theoretically
\cite{EGGK95,EK06,EKKAG09} and experimentally
\cite{HACCES06,WJF07,JWF07}. Similar approach appeared to be very
successful in studying temporal dynamics of tunneling
\cite{DFSS07,DFSF09,DFFS10}. This analogy prompted a proposal
\cite{FFBF10} to create an optical analog of the Laval nozzle, in
which the luminous fluid is accelerated to a supersonic velocity
and a sonic horizon is created. Our recent experiment \cite{EFB12},
realizing the proposal \cite{FFBF10}, demonstrated a possibility of
transonic acceleration of a luminous fluid in an all-optical
analog of the Laval nozzle. A laser beam is propagating in a
specially profiled cavity filled with ethanol with an addition of
iodine. The beam disperses and also deviates from the straight line
(along the $z$ axis) propagation so that the beam angle with respect
to the axis mimics the flow velocity and its variation mimics
acceleration. The defocusing nonlinearity in this experiment appears
due to a local temperature variation under the action of the laser
beam. Such mechanism directly leads to nonlocality of the Kerr
nonlinearity (see review \cite{KBNNWRE04}). It was demonstrated both theoretically and
experimentally in Ref. \onlinecite{MNDKK07} that the nonlocality can
be described by a nonlocal response function with a finite radius.
The nonlocality in Kerr nonlinear media allows one to control
stability and mobility of solitons (bright or dark, depending on the
sign of the nonlinearity) as shown theoretically \cite{KVT04,YKT08,NNKBRC04} and confirmed experimentally. \cite{DNPBK06} In
Bose-Einstein condensates the nonlocality may become of importance
at high enough densities \cite{PSR98}. It is also typical of the
dipolar bosonic quantum gases \cite{LMSLP09} and of nematic liquids \cite{RBK05}.

However the issue of nonlocal nonlinearity has not yet been
addressed in the context of the horizon physics. A number of papers
\cite{CFRBF08,RPC09,MP09a,MP09b,FP11,FS11,LRCP12,KP12} present
detailed analysis of fluctuations and Hawking radiation in
transonically accelerated fluid accounting for local nonlinearity
only, which is a quite reasonable approximation in BEC at least for
very short-range interactions. The same approximation in Kerr
nonlinear media may be less harmless and an analysis of the role of
nonlocal nonlinearity in dynamics of fluctuations and formation of
Hawking radiation near the Mach horizon is necessary. The
nonlocality introduces a new length $\sigma$ (see Ref. \onlinecite{MNDKK07}) so that
we may expect that fluctuations with shorter lengths will be smeared
out. The interplay of the nonlocality length $\sigma$ with other lengths characterizing
the problem will be discussed in what follows and conditions for
observation of analog Hawking radiation in nonlocal Kerr nonlinear
media will be formulated.

\section{Nonlocal nonlinearity.}

We consider here the Nonlinear Schr\"odinger (NLS) equation in $1 +
1$ dimensions,
\begin{equation}\label{NLS}
i \partial_z \Psi(x, z) =
$$$$
- \frac{1}{2\beta} \partial_x^2 \Psi(x, z) + g (\widehat{R}
|\Psi|^2)(x,z) \Psi(x, z)
\end{equation}
where $\Psi$ is a paraxial amplitude of light propagating along the
$z$ axis, which plays the part of time. $\beta$ is the wave vector
of the laser beam, which plays now the role of mass. $\widehat{R}$
is a linear integral nonlocal operator,
\begin{equation}\label{kernel-a}
\widehat{R} |\Psi|^2 = \int dx'dz' R(x'-x ,z'-z ) |\Psi(x',z')|^2
\end{equation}
The kernel of this operator is normalized to one and is assumed to be not singular and characterized by a finite length scale $\sigma$. For example it may be chosen in the form
\begin{equation}\label{kernel-b}
R(x,z) = \frac{1}{2\pi\sigma^2} e^{- \sqrt{x^2 + z^2}/\sigma}.
\end{equation}
In Kerr nonlinear optics the nonlocality depends obviously both on
$x$ ("coordinate") and $z$ ("time").  The dependence is expected to
be symmetric. The Fourier transform of (\ref{kernel-b}) reads
$$
\widetilde{R}(k_x,k_z) = \frac{1}{(1 + \sigma^2k_x^2+ \sigma^2 k_z^2)^{3/2}}.
$$
It tends to one in the long wave limit $k_x,k_z \to 0$, and to zero in the short wave limit $k_x,k_z \to \infty$. These generic features (see, e.g. discussion in Ref. \onlinecite{WKBR02}) will be important for the analysis to be presented in the following.

The Madelung transformation $\Psi = f e^{-i \varphi}$ allows one to
represent the NLS equation for a complex function as two
hydrodynamic equations
\begin{equation}\label{continuity}
\partial_z \rho + \partial_x (\rho v) = 0,
\end{equation}
\begin{equation}\label{Euler}
\partial_z v + \frac{1}{2}\partial_x v^2 = -
\frac{1}{\beta} \partial_x \left[U_{qu} + g \widehat{R} \rho
\right]
\end{equation}
for two real functions. Here
$$
U_{qu} = - \frac{1}{2\beta}\frac{\partial_x^2 f}{f}.
$$
These two equations describe an equivalent luminous fluid, where
the light intensity $\rho = f^2$ plays the role of the density and
$\beta v = - \partial_x \varphi$ defines the flow velocity $v$ in
the $x$ direction.

Our aim here will be to analyze the behavior of fluctuations near
the Mach horizon of the transonically accelerating luminous fluid.
This analysis will be quite analogous to the one described in detail
in Ref. \onlinecite{FS11} (references to some earlier papers can
also be found there). The only difference is due to the nonlocality
term in the Euler equation (\ref{Euler}). That is why only principal
steps will be outlined below, which are necessary to introduce the
notations and arrive at the result.

We consider fluctuations on the background of a given stationary
flow profile described by a function $\Psi_0 = f_0 e^{-i\varphi_0}$. They can appear spontaneously or induced artificially by experimental means.\cite{FFBF10}
Linearized equations for these fluctuations are deduced from
(\ref{continuity}) and (\ref{Euler}),
\begin{equation}\label{linear-b}
\begin{array}{c}
(\partial_z + v_0 \partial_x) \chi - \displaystyle \frac{1}{\beta \rho_0}
\partial_x(\rho_0 \partial_x \xi) = 0,
\\ \\
(\partial_z + v_0 \partial_x) \xi + \frac{1}{4\beta\rho_0}
\partial_x( \rho_0 \partial_x \chi) - g(\widehat{R}\rho_0\chi) =0.
\end{array}
\end{equation}
where the density $\delta\rho(x,z)$ and velocity $\delta v(x,z)$
fluctuations are defined by the relations
\begin{equation}\label{fluctuations-c}
\begin{array}{c}
\delta v(x,z) = - \displaystyle \frac{1}{\beta}
\partial_x\xi(x,z),
\\
\delta\rho(x,z) = \rho_0(x) \chi(x,z),
\end{array}
\end{equation}
and $\rho_0 = f_0^2$, $v_0 = - (1/\beta) \partial_x \varphi_0$.

Equations resulting from linearization around a known solution are usually called modulation equations, which may or may not lead to a modulation instability. Such an instability appears typically in the case of negative $g < 0$, i.e. focusing nonlinearity, whereas we deal here with positive, $g > 0$, i.e. defocusing nonlinearity. Refs. \onlinecite{KBRW01,WKBR02} present a general discussion in the context of the Kerr nonlinear media. In the simple case of a constant background density $\rho_0$ and
velocity $v_0$ we readily obtain the spectrum of the fluctuations in
the form
\begin{equation}\label{Bogoliubov}
(k_z - v_0 k_x)^2 =\frac{g\rho_0}{\beta} k_x^2
\left[\frac{k_x^2l_n^2}{2} + \widetilde{R}(k_x,k_z) \right],
\end{equation}
which corresponds to the Bogoliubov excitation spectrum in the case
of nonlocal interaction. Here $l_n^2 = 1/(2 \beta g\rho_0)$ is the
nonlinearity length (healing length in BEC). If the nonlocality
kernel is chosen in the form (\ref{kernel-b}) then the long wave limit in (\ref{Bogoliubov}) holds at $k_{x} \sigma \ll 1$ and $k_{z} \sigma \ll 1$ and the sound velocity keeps its
standard form $s^2 = g\rho_0/\beta $. We assume the nontrivial
situation when $\sigma > l_n$. Then the quartic dependence
$$
(k_z - v_0 k_x)^2 = \frac{k_x^4}{4 \beta^2}
$$
becomes dominant under the condition that $k_{x,z} > (l_n^2
\sigma^3)^{-1/5}$, which may be fulfilled even at $l_n k_{x,z} < 1$.

\section{Regularization due to the nonlocality.}

In order to consider the fluctuations close to the Mach horizon we
will use the relations
$$
v = \overline{s}(1 + \alpha x),\ \ \frac{g\rho_0(x)}{\beta} =
\overline{s}^2 (1 - \alpha x).
$$
describing a spatially accelerating flow. Here $\overline{s}$ is the
sound velocity at $x=0$, i.e. at the throat of the Laval nozzle.
Then Eqs. (\ref{linear-b}) take the form
\begin{widetext}
\begin{equation}\label{2}
\begin{array}{c}
\partial_z\chi + \overline{s}(1 + \alpha x)
\partial_x \chi + \displaystyle\frac{1 }{\beta} \left[\alpha
\partial_x  - \partial^2_x \right] \xi = 0,
\\ \\
\displaystyle\frac{1}{4\beta} \left[- \alpha \partial_x +
\partial_x^2\right] \chi - \beta \overline{s}^2 \widehat{R} \chi +
\beta \overline{s}^2 \alpha \widehat{R} ( x \chi) + [\partial_z +
\overline{s}(1 + \alpha x) \partial_x]\xi = 0.
\end{array}
\end{equation}
\end{widetext}
where the terms $ O(\alpha^2 x^2)$ are omitted. These equations can
be solved in the Fourier space to within the terms $O(\alpha/k_x)$
(see Ref. \onlinecite{FS11}). The Fourier transformed first equation
(\ref{2}) is solved with respect to $\xi_k$, which is then
substituted into the second equation. As a result we get
\begin{widetext}
\begin{equation}\label{difequ-a}
\partial_{k_x} \ln \chi_k \approx i
\frac{ \displaystyle\frac{l_n^2}{2} (2i\alpha k^3_x + k_x^4) +
(i\alpha k_x + k_x^2) \widetilde{R}(k_x,\nu) - (\nu - k_x)^2 - i\alpha
k_x}{\alpha k_x \{2\nu - [2 + \widetilde{R}(k_x,\nu)] k_x - i\alpha
\widetilde{R}(k_x,\nu)\}}
\end{equation}
\end{widetext}
where $\nu = k_z/\overline{s}$.

The integration of the r.h.s. of Eq. (\ref{difequ-a}), although
possible, may result in very cumbersome expression. That is why we
consider here two limits. If $\sigma k_{x,z} \ll 1$, then
$\widetilde{R}(k_x,\nu) \approx 1$ and we get
\begin{equation}\label{difequ-d}
\partial_{k_x} \ln \chi_k(k_x,\nu) =
$$$$
i \frac{ \displaystyle \frac{l_n^2}{2} (2i\alpha k_x^3 + k_x^4) -
(\nu^2 - 2\nu k_x)}{\alpha k_x (2\nu - 3k_x - i\alpha )}.
\end{equation}
It means that we return to the situation of the local nonlinearity.
As shown in Ref. \onlinecite{FS11} there is a singular real space
solution of Eq. (\ref{2})
\begin{equation}\label{singular}
\chi_s(x,\nu) \propto x^{\gamma - 1},
\end{equation}
where
\begin{equation}\label{exponent}
\gamma = \frac{2i\nu}{3 \alpha} + \frac{4i}{81} \frac{l_n^2
\nu^3}{\alpha} - \frac{2}{27} l_n^2 \nu^2.
\end{equation}
Finally it results in the $\nu$-dependent Hawking temperature
$$
T_H(\nu) = T_H(0)/(1+ \frac{2}{27} l_n^2 \nu^2)
$$
where
$$
T_H(0) = \frac{3\hbar \overline{s}}{4\pi k_B}.
$$
This solution holds for the distances from the Mach horizon
satisfying the condition $\min\{1/\nu, 1/\alpha\} \gg |x|$. In the
absence of nonlocality the other condition is $|x| \gg l_r$ where
$l_r = l_n/(l_n\alpha)^{1/3}$ is the regularization
length\cite{FS11} (see also\cite{FP11}). As we will see below the
nonlocality also leads to a regularization which means that the
final condition reads $|x| \gg \max\{\sigma, l_r\}$.

We also have to consider the limit $k_{x,z}\sigma \gg 1$. It
produces nontrivial results only if $\sigma > l_n$. Otherwise the
problem can be reduced to the local one. In the limit $k_x\sigma \gg
1$ when $\widetilde{R}(k_x,\nu) \ll 1$ equation (\ref{difequ-a}) becomes
\begin{equation}\label{difequ-e}
\partial_{k_x} \ln \chi_k(k_x,\nu) =
$$$$
i \frac{\displaystyle\frac{l_n^2}{2} (2i\alpha k_x^3 + k_x^4) -
\nu^2 + 2\nu k_x - k_x^2 - i\alpha k_x}{2\alpha k_x (\nu - k_x )}.
\end{equation}
Carrying out the procedure, as outlined in Ref. \onlinecite{FS11},
the real space density fluctuations for a given $\nu$ are described
by the function
\begin{equation}\label{fluctuations-d}
\chi_\nu(x,z) = e^{-ik_zz}\int dk k^{\gamma'_1} (k -
\nu)^{\gamma'_2} e^{ikx + \Lambda}
\end{equation}
where
$$
\begin{array}{c}
\displaystyle \gamma'_1 = - \frac{i \nu}{2\alpha},\\
\\
\displaystyle \gamma'_2 = - \frac{1}{2} + \frac{1}{2} \nu^2 l_n^2 -
\frac{i}{4\alpha} l_n^2 \nu^3
\end{array}
$$
and
$$
\Lambda = 
$$$$
\frac{i}{2\alpha} k + \frac{1}{2} k l_n^2 \nu -
\frac{i}{4\alpha} l_n^2 \nu^2 k + \frac{1}{4} l_n^2 k^2 -
\frac{i}{8\alpha} k^2 l_n^2 \nu - \frac{i}{12\alpha} l_n^2 k^3.
$$

The behavior of the density fluctuations can be readily deduced from
(\ref{fluctuations-d}). At $\nu \ll l_n^{-1}$ and $ \sigma \gg |x| \gg
l_n$ we get
\begin{widetext}
\begin{equation}\label{fluctuations-e}
\chi_\nu(x,z) = e^{-ik_zz} \int dk k^{\gamma'_1} \left(k - \nu
\right)^{\gamma'_2} e^{ikx + \frac{i}{2 \alpha} k} \approx e^{-ik_zz} \left(x +
\frac{1}{2 \alpha} \right)^{-(\gamma'_1 + \gamma'_2 + 1)} \int dy
y^{\gamma_1 + \gamma_2} e^{iy}
\end{equation}
\end{widetext}
where integration variable $y = [x + 1/(2 \alpha)] k$ is used. It
means that the function $\chi$ tends to a constant (since $\alpha
|x| \ll 1$) and the singular solution (\ref{singular}) becomes
regular at small enough $x$. This regularization is not violated at
even smaller distances $|x| \ll l_n$ and may be obtained in the way
similar to that used in Ref. \onlinecite{FS11}.

\section{Discussion and summary.}

As discussed in Introduction, the non-locality of the nonlinearity
may play an important part in experiments designed to observe
Hawking-type classical and quantum fluctuations near the Mach
horizon formed by a transonic flow of luminous fluid. The above
derivation shows that the non-locality should be taken into account
if its characteristic length scale exceeds the regularization
length, $\sigma > l_r$. The largest of these determines the point,
below which the singular fluctuation eigenfunction (\ref{singular})
is regularized. Consequently, if $\sigma > l_r$ it is the length
scale $\sigma$, characterizing the nonlocality, which sets the lower
bound for the distance $|x|$ such that Hawking radiation or
straddled fluctuations appear.

\begin{figure}[htb]
\includegraphics[width=0.5\textwidth,angle = 0]{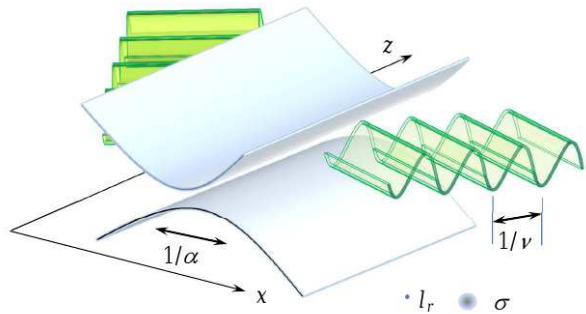}
\caption{(Color online) A sketch of an all-optical Laval nozzle
demonstrating the relative length scales which determine the window
in real space where straddled fluctuations may be generated in an
experiment. $\sigma$ is the non-locality length (typically 1 mm);
$l_r$ is the regularization length, determined by the nonlinear
length and the acceleration rate (typically a few tenths of a mm);
$\alpha$ is the acceleration rate (typically $ \geq 1 mm^{-1}$);
$\nu$ is the characteristic "frequency" of the fluctuations, and
should be such that $1/\nu$ be of the order of the size of the
experimental test cell (a few centimeters).} \label{Sketch2}
\end{figure}

At the same time it should be taken into account that the derivation
of (\ref{singular}) holds for distances from the horizon that are
not too large, and in particular shorter than $1/\alpha$ (the
inverse acceleration rate) and $1/\nu$ (the length scale of a
fluctuation in the $z$ direction). The latter sets an upper limit
for the distance from the horizon where a fluctuation may develop
(see e.g. Ref. \onlinecite{FFBF10} for a discussion of a Kerr nonlinear
medium). A window $\min\{1/\nu, 1/\alpha\} >  |x| > \max\{l_r,
\sigma\}$ in real space must therefore exist in order for
fluctuations to be observed experimentally. In a typical experiment,
such as the one described in \cite{EFB12}, the nonlinear length is on
the order of several tens of microns, while the non-locality length
is on the order of 1 mm \cite{MNDKK07}. This means that $1/\nu$ and
$1/\alpha$ must both be at least on the order of a few millimeters.
$1/\nu$ can be as large as the size of the experimental apparatus (a
few centimeters), and is not expected to be the limiting constraint.
On the other hand, the experimental setup must be designed in such a
way that the acceleration of the luminous fluid should occur over a
distance of several millimeters in order that the required window
for the fluctuations exists. Changing the stationary profile allows
one to change the spatial acceleration $\alpha$ and therefore $l_r$.
This offers the possibility by increasing $\alpha$ to turn to the
regime where $\sigma$ is the length scale separating the regular
from the singular behavior for all $\alpha > \alpha_c =
(l_n/\sigma)^2/\sigma$. In case when the quantum potential $U_{qu}$
can be neglected, i.e. we can assume that $l_n =0$, it is the
nonlocality regularizing the fluctuations for $|x| < \sigma$. The
relationships between these different length scales are
schematically shown in Fig. \ref{Sketch2}.

To summarize, we considered the effect of the non\-locality of the
nonlinear response on the feasibility of observing, in an actual
experiment, fluctuations near a "sonic" horizon of an accelerating
luminous fluid that are analogs of Hawking radiation. We find
that the non-locality sets a lower bound on the distance from the
horizon where the fluctuations may develop, and this, in turn, sets
an upper bound on the acceleration rate of the fluid. With proper design of the experimental apparatus, there
appears to be a large enough window for observing such fluctuations.

{\bf Acknowledgment} SB and VF acknowledge the financial
support of the Israeli Science Foundation, grant N 1309/11. VF acknowledges
fruitful discussions with G. Shlyapnikov and E.
Bogomolny during his visit to LPTMS,Univ. Paris-Sud, Orsay.

\end{document}